\def\norm[#1]{\lVert{#1} \rVert}
\def\speedOfLight{c}
\def\distance{R_{\indexAP\indexLED}}
\def\Ar{A}
\def\FOV{\theta_{\rm FOV}}
\def\mode{\gamma}
\def\angleLOSandTX{\phi_{\indexAP\indexLED}}
\def\angleLOSandRX{\theta_{\indexAP\indexLED}}
\def\rect[#1]{{\Pi}\left(#1\right)}
\def\dirac[#1]{{\delta}\left(#1\right)}
\def\realNumbers{\mathbb{R}}
\def\location[#1]{{\rm \bf r}_{#1}}
\def\orientation[#1]{{\rm \bf q}_{#1} }
\def\channell[#1]{h(#1)}
\def\source{\mathcal{S}_{\indexAP\indexLED}}
\def\receiver{\mathcal{R}}

\def\propagationDelay{\tau}

\def\indexUser{R}
\def\indexLED{n}
\def\indexAP{k}
\def\LEDGroup[#1]{l(#1)}
\def\LEDGroupNumber[#1]{l_{#1}}
\def\LEDConnectFlag[#1][#2]{\alpha_{#1#2}}
\def\channel[#1][#2]{h_{#1#2}}
\def\channelVector[#1][#2]{\textbf{h}_{#1#2}}
\def\power[#1]{p_{#1}}

\def\ReceivedPower[#1][#2][#3]{H_{#3,#1(#2)}}
\def\signalVector[#1][#2]{\textbf{v}_{#2#1}}

\documentclass[journal]{IEEEtran}
\renewcommand{\IEEEbibitemsep}{0pt plus 0.5pt}
\makeatletter
\IEEEtriggercmd{\reset@font\normalfont\fontsize{7.9pt}{8.40pt}\selectfont}
\makeatother
\IEEEtriggeratref{1}

\usepackage{array}
\usepackage[cmex10]{amsmath}
\usepackage{amssymb}
\usepackage{algorithmic}
\usepackage{acronym}
\usepackage{graphicx}
\usepackage{float}
\usepackage[noadjust]{cite}
\usepackage{etoolbox}%
\usepackage{amsthm}%
\usepackage{xcolor}
\usepackage{multirow}
\usepackage{subfigure}

\theoremstyle{remark}

\begin{document}

\title{Adaptive Kalman Tracking for Indoor Visible Light Positioning}

\author{\IEEEauthorblockN{Yusuf Said Eroglu, Fatih Erden, and Ismail Guvenc}\\
\IEEEauthorblockA{Department of Electrical and Computer Engineering, North Carolina State University, Raleigh, NC\\
Email: \{yeroglu, ferden, iguvenc\}@ncsu.edu}
\thanks{This work is supported in part by NSF CNS award 1422062.}
}
\vspace{-2cm}
\maketitle
\begin{abstract}
Visible light communication~(VLC) utilizes light-emitting diodes~(LEDs) to transmit wireless data. A VLC network can also be used to localize mobile users in indoor environments, where the global positioning system~(GPS) signals are weak. However, the line-of-sight (LOS) links of mobile VLC devices can be blocked easily, which decreases the accuracy of localization. In this paper, we study tracking a VLC user when the availability of VLC access point~(AP) link changes over the user's route. We propose a localization method for a single available AP and use known estimation methods when a larger number of APs are available. Tracking mobile users with Kalman filter can increase the accuracy of the positioning, but the generic Kalman filter does not consider instant changes in the measurement method. In order to include this information in the position estimation, we implement an adaptive Kalman filter by modifying the filter parameters based on the availability of APs to the user. Simulation results show that the implemented method decreases the root-mean-square error~(RMSE) of the localization down to 30\%-50\% of the original estimation. 
\end{abstract}


\section{Introduction}

There is an increasing demand for indoor positioning and navigation systems due to the proliferation of location-based services. The global positioning system (GPS) does not work well in indoor environments because of the strong signal attenuation. Dedicated indoor localization systems can help to provide accurate location information needed for these services at the expense of deployment and operation costs. Several indoor positioning systems have been proposed using different wireless technologies, including radio frequency identification (RFID), infrared (IR), and wireless local area network (WLAN). \looseness=-1

Another alternative method for indoor localization is visible light positioning (VLP), which has lately received attention~\cite{6685759, yin2015indoor, Alphan_JLT}. The energy-efficient light-emitting diodes (LEDs) are replacing older light sources such as incandescent or fluorescent bulbs. Besides energy efficiency, LEDs can conveniently offer many new applications such as visible light communication (VLC) and positioning. VLC provides an additional wireless communication spectrum which is wide and unregulated and can complement radio frequency (RF) on providing wireless connectivity. Since visible light does not penetrate walls, the inherent physical security makes VLC a good candidate for military applications. Other advantages of the VLC are the energy efficiency due to using the same signal for illumination and communication, the absence of radio interference, and high spectral reuse~\cite{6655152}. As VLC starts to serve users for communication purposes, an inherent application would be to use it for indoor positioning. VLP can provide very accurate localization with a root-mean-square error (RMSE) below 5~cm when a sufficient number of LEDs are available to the user~\cite{Alphan_JLT}. \looseness=-1

VLP accuracy may suffer from the insufficient number of visible light access points (APs) or LEDs. To enhance the accuracy of the VLP, Kalman filters (KFs) can be used~\cite{eroglu_2015, 6477760, 6776093, 7925652}. The KF predicts the user location based on previous locations and combines the predictions with the actual measurements to cancel out the negative effects of instantaneous bad measurements. It can track the user efficiently in case the localization method itself fails to do so. However, in many applications, the estimation methods and/or parameters change frequently. To incorporate these changes into the KF, adaptive estimation can be used~\cite{Wang_1999}. The main advantage of the adaptive technique is its weaker reliance on the a priori statistical information. An adaptive filter formulation tackles the problem of imperfect a priori information and provides a significant improvement in performance over the fixed filter through the filter learning process~\cite{mehra1970}. It has been used for improving the accuracy of GPS \cite{mohamed1999adaptive}, and for tracking mobile Wi-Fi users with intermittent link availability~\cite{8049459}. \looseness=-1

The VLC signal highly depends on line-of-sight (LOS) link, and when the LOS links are blocked or misaligned~\cite{eroglu_TCOM}, some of the APs may not be available to the user for some time. The varying number of available APs requires the device to use different localization methods with different expected accuracies over the route. Therefore, the implementation of an adaptive KF is crucial for efficient tracking of the VLC users, which have not been addressed in the literature. In this paper, we implement an adaptive KF for VLC user tracking while the number of accessible APs varies with the user position. We propose a localization method for the case with a single available AP and adapt existing techniques with adaptive KF framework for the cases with multiple available APs. We assume each AP can be blocked with a random probability, and decide certain estimation models based on the resulting layout of available APs. We propose a heuristic method to weight each layout model according to their expected estimation accuracy and use these in the adaptive filter. We track the user over a random waypoint (RWP) route with conventional and adaptive KFs. The results show that the adaptive filter provides a significant increase in the localization accuracy compared to the conventional filter. \looseness=-1

The rest of this paper is organized as follows. Section~\ref{sec:system_model} presents the channel model, AP architectures with different LED configurations, and AP layout models considered in the study. Section~\ref{sec:localization} introduces the localization methods for different number of APs. Section~\ref{sec:tracking} describes the KF-based VLC user tracking. Section~\ref{sec:results} presents the simulation results, and Section~\ref{sec:conclusion} concludes the paper.

\textit{Notations:} $\textbf{I}$ is the identity matrix, $\rm{diag}(\textit{x})$ is the diagonal matrix with diagonal entries $x$.The transpose operation is denoted by $(\cdot)^T$. The Gaussian distribution with zero-mean and covariance matrix $\textbf{C}$ is shown as $\mathcal{N}(0, \textbf{C})$, the uniform distribution between $a$ and $b$ is shown as $\mathcal{U}(a, b)$, and the Euclidean norm is shown as $\| \cdot \|$.

\begin{figure}[tb]
	\centering
	\includegraphics[width = 2.9 in]{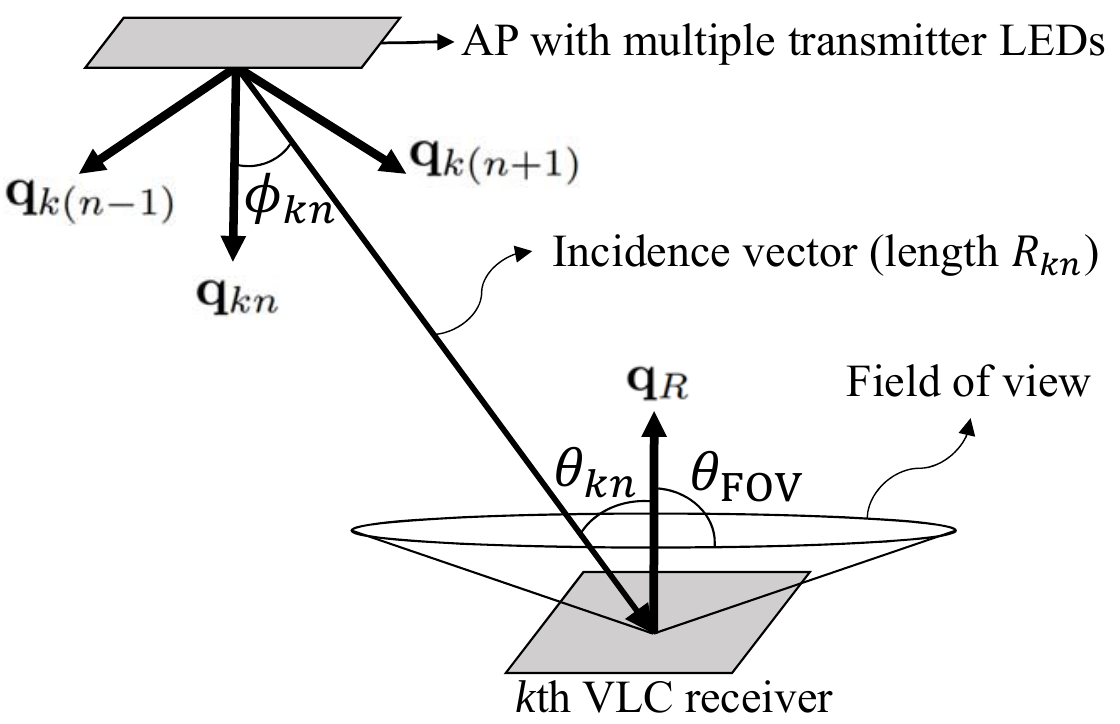}
	\caption{\small The LOS channel model with an AP and a receiver. }
	\label{LambertianFigure}
	\vspace{-1mm}
\end{figure}

\section{VLP System Model}\label{sec:system_model}
\subsection{Optical Channel Model}
We consider a scenario with multiple APs that communicate with a VLC receiver. Each AP has multiple co-located LEDs with different orientations as illustrated in Fig.~\ref{LambertianFigure}. The $\indexLED$th LED that is located on the $k$th AP, without loss of generality, can be defined with three parameters as $\source~=~\{\location[\indexAP\indexLED],\orientation[\indexAP\indexLED],\mode \},$ where $\location[\indexAP\indexLED] = [r_{\indexAP\indexLED}^{(\rm x)}, r_{\indexAP\indexLED}^{(\rm y)}, r_{\indexAP\indexLED}^{(\rm z)}]^{\rm T} \in\realNumbers^{3\times1}$ is the location of the LED,  $\orientation[\indexAP\indexLED] = [q_{\indexAP\indexLED}^{(\rm x)}, q_{\indexAP\indexLED}^{(\rm y)}, q_{\indexAP\indexLED}^{(\rm z)}]^{\rm T} \in\realNumbers^{3\times1}$ is the orientation of the LED, and $\mode$ is the parameter that specifies the directionality of the light source based on Lambertian pattern~\cite{Barry_1993}. Similarly, the receiver is modeled as $\receiver~=~\{\location[\indexUser],\orientation[\indexUser],\Ar, \FOV \},$ where $\location[\indexUser]\in\realNumbers^{3\times1}$ is the location of the photo-diode (PD), $\orientation[\indexUser]\in\realNumbers^{3\times1}$ is the orientation of the {PD}, $\Ar$ is the area of the PD in ${\rm cm}^2$, and $\FOV$ is the semi-angle field of view~(FOV) of the PD. All orientation vectors are unit vectors.

The {LOS} component of the channel impulse response between the source $\source$ and the receiver $\receiver$ is given by \cite{Barry_1993}
\begin{align}
h_{\indexAP\indexLED}(t; \source, \receiver)&=\frac{\mode+1}{2\pi}\cos^\mode(\angleLOSandTX) \cos(\angleLOSandRX)\frac{\Ar}{\distance^2} \nonumber \\&\times\rect[{\frac{\angleLOSandRX}{\FOV}}]
\rect[{\frac{\angleLOSandTX}{\pi/2}}]\dirac[t-\propagationDelay]
~,
\label{eq:LOSchannel}
\end{align}
where $\angleLOSandTX$ is the angle between $\orientation[\indexAP\indexLED]$ and the incidence vector, which is the vector from the transmitter LED to the receiver PD, and $\angleLOSandRX$ is the angle between $\orientation[\indexUser]$ and the incidence vector. $\distance$ is the length of the incidence vector, which is given as $\|{\location[\indexAP\indexLED] - \location[\indexUser]}\|$. The propagation delay is given as $\propagationDelay=\distance/\speedOfLight$, where $\speedOfLight$ is the speed of light. $\dirac[\cdot]$ denotes the Dirac function, and $\rect[\cdot]$ denotes the rectangle function which takes the value 1 for $|x| \le 1$, and 0 otherwise.

While $\rect[{{\angleLOSandRX}/{\FOV}}]$ in \eqref{eq:LOSchannel} implies that the receiver can detect the light only when $\angleLOSandRX$ is less than $\FOV$, $\rect[{{\angleLOSandTX}/{(\pi/2)}}]$ ensures that the location of the receiver is in the {FOV} of the source. The terms in \eqref{eq:LOSchannel} can be obtained as
\begin{align}
    \cos(\angleLOSandTX)=\orientation[\indexAP\indexLED]^{\rm T}({\location[\indexUser] - \location[\indexAP\indexLED]})/\distance,
\end{align}
and 
\begin{align}
    \cos(\angleLOSandRX)=-\orientation[\indexUser]^{\rm T}({\location[\indexUser] - \location[\indexAP\indexLED] } )/\distance \,.
\end{align}

\begin{figure*}[tbh]
\centering
\subfigure[3-LED AP architecture]{
   \includegraphics[height = 1.4in] {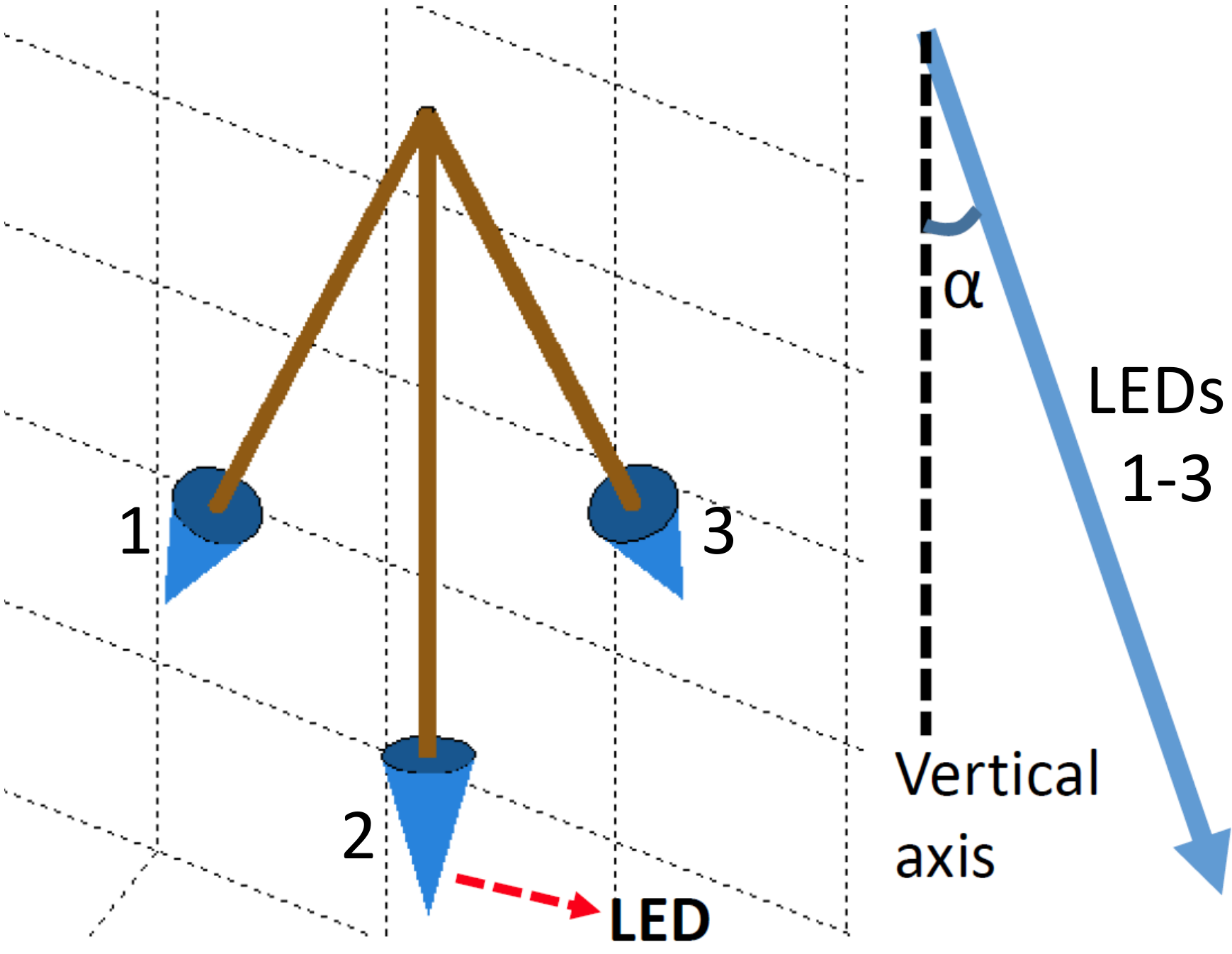}
   \label{3Element}
 }
 \quad
 \subfigure[7-LED AP architecture]{
   \includegraphics[height = 1.4in] {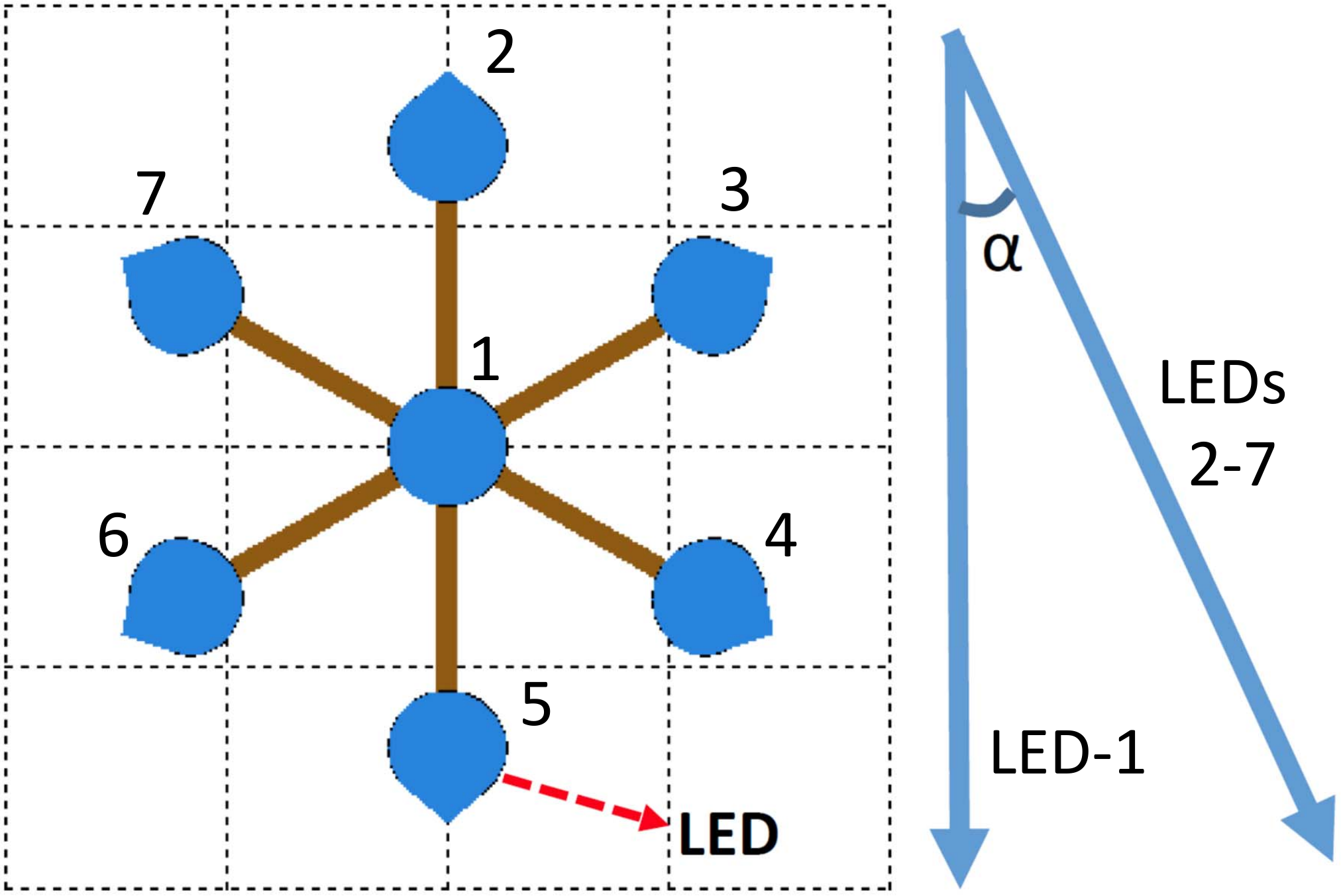}
   \label{7Element}
 }
 \quad
 \subfigure[19-LED AP architecture]{
   \includegraphics[height = 1.4in] {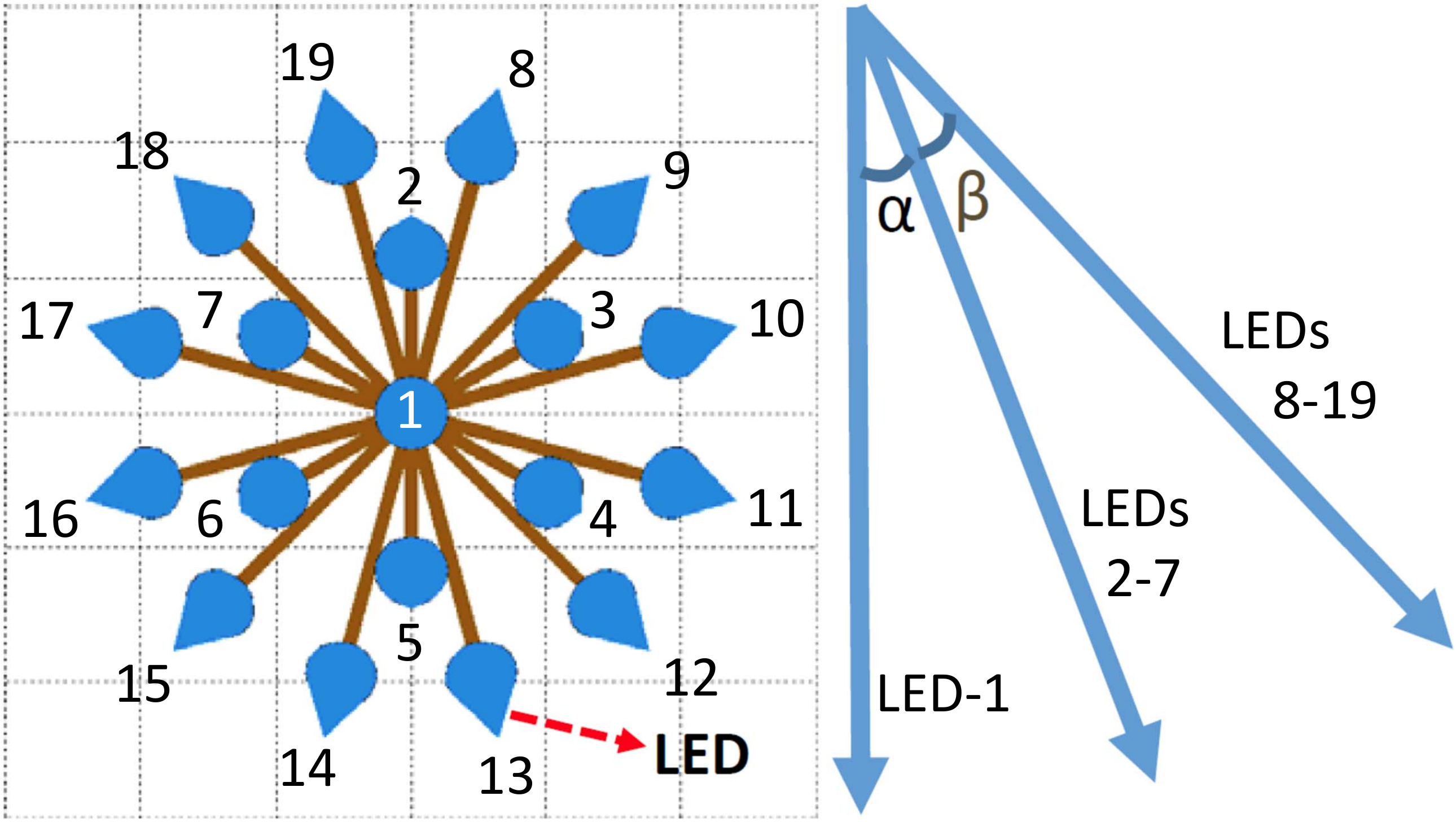}
   \label{19Element}
 }
\caption{Side views of three example multi-LED AP architectures that are considered for the simulations: (a) 3-element, (b) 7-element, and (c) 19-element transmitter. All transmitters are mounted at the corners of the room at the ceiling height, facing the center of the room with a $45^\circ$ separation from each wall.}
\label{APs}
\vspace{-1mm}
\end{figure*}

\subsection{AP Architectures}
This subsection explains the AP architectures considered for the simulations in this manuscript. We consider APs with multiple co-located LEDs that have different orientations as examined in~\cite{chen2015space, eroglu_GlobecomWS, Eroglu_JSAC, yin2015indoor}. These angle diversity APs provide advantages for communication purposes such as space division~\cite{chen2015space, eroglu_GlobecomWS}, and localization purposes such as better angular resolution \cite{Eroglu_JSAC}. 

Some of the example AP architectures are illustrated in Fig.~\ref{APs}. Fig.~\ref{3Element} shows the AP with three LEDs. The LEDs are vertically tilted with $\alpha$ degree from the center orientation of the AP. For APs with four to seven LEDs, we use the architecture in Fig.~\ref{7Element}, where one LED is in the middle, and other LEDs surround it with again $\alpha$ degree vertical tilt, and they are separated from each other with equal degrees on the horizontal plane. For APs with more than seven LEDs, LED configuration in Fig.~\ref{19Element} is used, where there is a single LED in the middle, six LEDs surround it in the first layer with $\alpha$ degree vertical tilt, and the rest of the LEDs surround them in the second layer with an additional $\beta$ degree vertical tilt. Again, the LEDs on any layer are separated from each other with equal degrees, except for the AP with eight LEDs where there is a single LED on the second layer.

\subsection{AP Layout Models} \label{LocModels}
Without loss of generality, we consider a square room with four APs located at the corners of the room at the ceiling height. The APs are oriented so that they face the center of the room, with 45$^\circ$ separation from all three walls. Each AP has multiple co-located LEDs with different orientations, as shown in Fig.~\ref{APs}. As the user moves around the room, some of the APs might be blocked due to intervening objects, change in the receiver orientation~\cite{eroglu_TCOM}, or the limited FOV of the PD. The receiver then localizes itself using the signal that it receives from all the available APs. 

\begin{figure}[tb]
	\centering
	\includegraphics[width = 3.3 in]{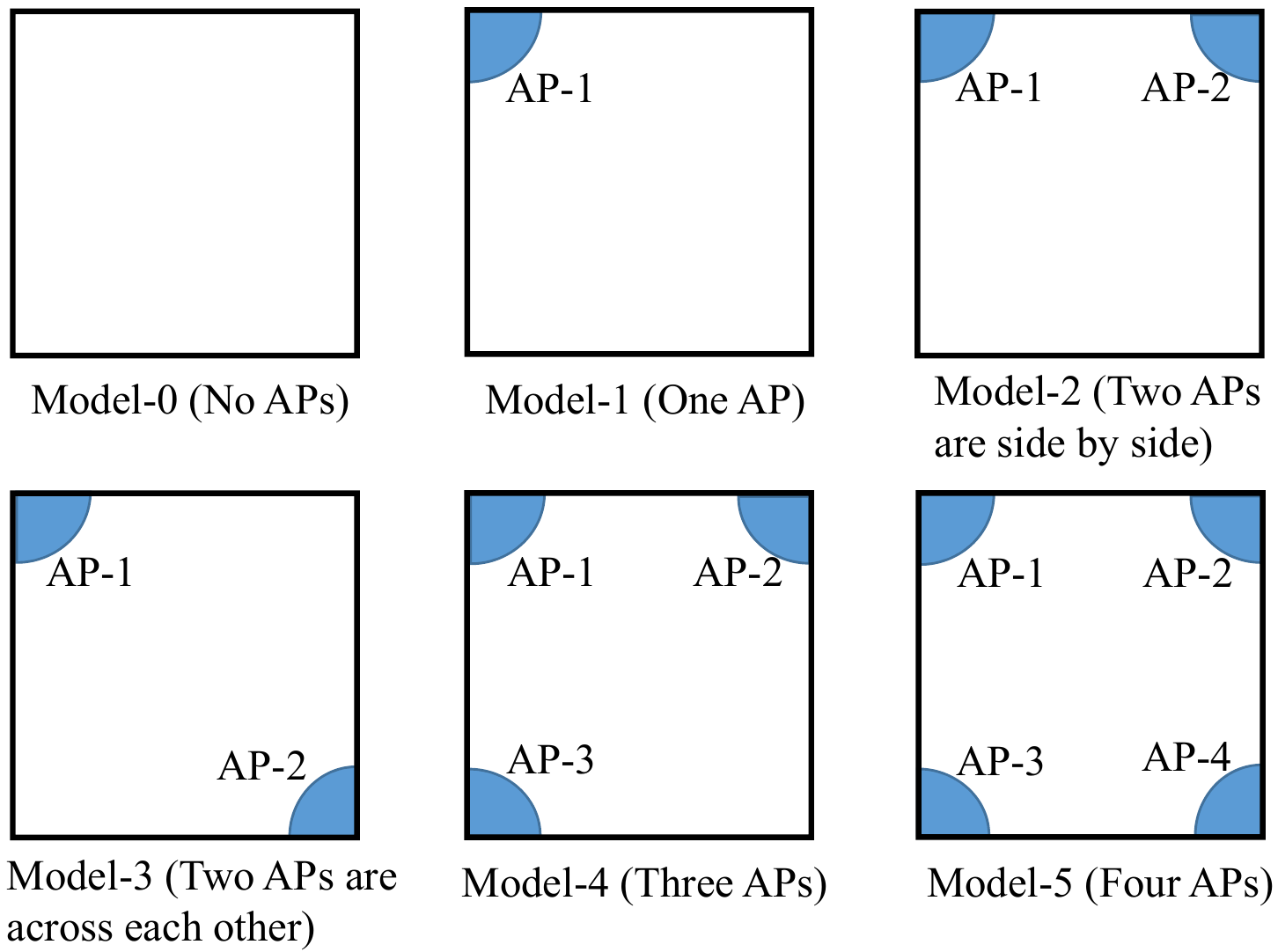}
	\caption{\small AP layouts based on the availability of the APs. Blue corners in the models represent the APs that are available to the user, whereas empty corners represent the APs that are blocked or not in the FOV of the receiver.}
	\label{Models}
	\vspace{-1mm}
\end{figure}

As some of the APs are blocked, or not in the FOV, we can consider six different layout models for the accessible APs, as shown in Fig.~\ref{Models}. We refer to the layout with no available APs as Model-0. In this case, localization relies solely on the prediction from the previous location estimates. Model-1 refers to the layout where there is only one available AP. In this case, localization is performed as described in Section~\ref{SingleAP}. For other models, there are multiple available APs, and the hybrid localization method \cite{Alphan_JLT}, which is described in Section~\ref{moreAPs}, is used. 

\section{Localization Methods and AP Availability} \label{sec:localization}
In this section, we first describe the localization method for multiple available APs, and then for a single available AP. \looseness=-1

\subsection{Multiple APs} \label{moreAPs}
There are many localization methods that consider multiple APs \cite{Alphan_JLT, eroglu_2015, yin2015indoor}. In this paper, we use the hybrid localization method proposed in \cite{Alphan_JLT}. The method is named the hybrid localization because it uses both the angle-of-arrival (AOA) and the received signal strength (RSS) information. In this method, RSS information from the LEDs modeled as in~\eqref{eq:LOSchannel} are fed to a maximum likelihood~(ML) estimator. The iterative solution starts with an initial point, which is assumed to be the result of the AOA-based localization~\cite{eroglu_2015}. The hybrid localization provides high accuracy, especially when a large number of APs and LEDs are available.

\subsection{Single AP} \label{SingleAP}
When there is only a single AP available at a given time, neither ML estimation nor AOA-based localization works. In this case, we use the AOA information from the single AP to estimate the location of the receiver. In most indoor applications, the mobile receiver is either on a hand-held device or placed on a desk, and the AP is located higher than the receiver. By assuming a constant height for the receiver, and using the AOA information, the user location can be estimated by finding the intersection of a vector and a plane. Assume that the available AP is the $k$th AP, and the estimated AOA information from the $k$th AP is $\textbf{p}_k = [p_k^{(x)}, p_k^{(y)}, p_k^{(z)}]$, which is a unit vector that points in the direction from the AP to the receiver. The height of the receiver is denoted by $\nu$. Then, the estimated location of the user is the intersection of the vector $\textbf{p}_k$ and the horizontal plane defined by the height $\nu$, which is given as
\begin{align}
    \textbf{r}_{R}' = \textbf{r}_{\indexAP} - \textbf{p}_k(\nu - r_{\indexAP}^{(z)})/p_k^{(z)},
\end{align}
where 
$\textbf{r}_{\indexAP}$ is the location of the available AP, and $r_{\indexAP}^{(z)}$ is the height of it. The term $\textbf{p}_k$ can be estimated by a weighted average of the orientation vectors of the LEDs on the AP~\cite{Alphan_JLT}. The normalized AOA estimation vector can therefore be written as
\vspace{-1mm}
\begin{align}
    \textbf{p}_k = \frac{ \tilde{\textbf{p}}_k}{\| \tilde{\textbf{p}}_k \|}, ~ \textrm{where} \quad \tilde{\textbf{p}}_k = \sum_{n = 1}^N h_{\indexAP\indexLED}\textbf{q}_{\indexAP\indexLED},
\end{align}
where $h_{\indexAP\indexLED}$ is the signal strength received from the $n$th LED, and $N$ is the number of LEDs on the $k$th AP.

\section{Tracking the Mobile VLC User} \label{sec:tracking}
In this section, we present the conventional and the adaptive KF models, and techniques for tracking a mobile VLC user.

\subsection{Conventional Kalman Filter} \label{Kalman_Sub}
The KF is a Bayesian filter that can track a user by exploiting the prior location information. Let $(x_t, y_t, z_t)$ denote the location estimate of a VLC user at time $t$, and $(V_t^x, V_t^y, V_t^z)$ denote the velocity of the user in $(x, y, z)$ coordinates at time $t$. Then, the state vector containing the KF prediction for user location and velocity at time instant $t$ can be written as
\begin{align}
\!\underbrace{\left[\begin{array}{c} x_t \\ y_t \\ z_t \\ V_t^x \\ V_t^y \\ V_t^z \end{array} \right]}_{\hat{\textbf{x}}_{t}} \!=\! \underbrace{\begin{bmatrix} 1 & 0 & 0 & \Delta t & 0 & 0 \\ 0 & 1 & 0 & 0 & \Delta t & 0 \\ 0 & 0 & 1 & 0 & 0 & \Delta t \\ 0 & 0 & 0 & 1 & 0 & 0 \\ 0 & 0 & 0 & 0 & 1 & 0 \\ 0 & 0 & 0 & 0 & 0 & 1 \end{bmatrix}}_{\textbf{B}}\underbrace{\left[ \begin{array}{cc} x_{t-1} \\ y_{t-1} \\ z_{t-1} \\ V_{t-1}^x \\ V_{t-1}^y \\ V_{t-1}^z \end{array} \right]}_{\textbf{x}_{t-1}} \!+\! \underbrace{\left[ \begin{array}{cc} e^x \\ e^y \\ e^z \\ e^{V_x} \\ e^{V_y} \\ e^{V_z} \end{array} \right]}_{\textbf{e}_x}, \nonumber
\end{align}
where $\textbf{B}$ is the state transition matrix, $\Delta t$ is the time difference between two states, $\textbf{x}_{t-1}$ is the estimation from previous state, and $\textbf{e}_{\rm x} \sim \mathcal{N}(0, \textbf{E}_{\rm x})$ is the prediction error vector. Moreover, let $(z_t^x, z_t^y, z_t^z)$ denote the observation of the user at time step $t$ before applying the KF. Then, the observation equation can be written as
\begin{align}
\underbrace{\left[ \begin{array}{c} z_t^x \\ z_t^y \\ z_t^z \end{array} \right]}_{\textbf{z}_t^{\rm obs}}  =  \underbrace{\begin{bmatrix} 1 & 0 & 0 & 0 & 0 & 0 \\ 0 & 1 & 0 & 0 & 0 & 0 \\ 0 & 0 & 1 & 0 & 0 & 0 \end{bmatrix}}_{\textbf{C}} \underbrace{\left[ \begin{array}{c} x_t \\ y_t \\ z_t \\ V_t^x \\ V_t^y \\ V_t^z \end{array} \right]}_{\hat{\textbf{x}}_{t}} + \underbrace{\left[ \begin{array}{cc} v^x \\ v^y \\ v^z \end{array} \right]}_{\textbf{v}}~, \nonumber
\end{align}
where $\textbf{v} \sim \mathcal{N}(0, \textbf{E}_{\rm v})$ is the observation error. The term $\textbf{z}_t^{\rm obs}$ is only a temporal estimate. At time $t$, we also get measurement values for $x_t$, $y_t$, and $z_t$, and they form our measurement vector $\textbf{z}_t^{\rm meas}$. Then we calculate our final estimate using
\begin{align}
\textbf{x}_{t} = \hat{\textbf{x}}_{t} + \textbf{K}_t (\textbf{z}_t^{\rm meas} - \textbf{z}_t^{\rm obs})~, \label{kalman_gain}
\end{align}
where ${\bf K}_t$ is the Kalman gain. The Kalman gain is a matrix that provides a balance between the prediction from previous steps and the measurement from localization at the current step. It can be iteratively calculated at each step by
\begin{align}
&\textbf{P}_t = \textbf{BP}_t'\textbf{B}^T + \textbf{E}_{\rm x}, \\
&\textbf{K}_t = \textbf{P}_t\textbf{C}^T(\textbf{CP}_t\textbf{C}^T + \textbf{E}_{\rm v})^{-1}, \label{Ezz} \\
&\textbf{P}_{t+1}' = (\textbf{I} - \textbf{K}_t\textbf{C})\textbf{P}_t~, \label{cov_state}
\end{align}
where $\textbf{E}_{\rm x}$ is the constant covariance matrix of $\textbf{e}_{\rm x}$ and can be modeled as ${\rm diag}(\sigma_x^2)$ assuming the errors of the parameters are independent and of similar amplitude. The $\textbf{E}_{\rm v}$ is the covariance matrix of $\textbf{v}$ and can be modeled as ${\rm diag}(\sigma_{\rm v}^2)$, and $\textbf{P}$ is an error covariance matrix which is updated twice in an iteration. In the next subsection, we discuss the adaptive KF.

\subsection{Adaptive Filtering}
The conventional KF assumes that we do not have information about the accuracy of the measurement at each state, and it keeps track of the estimated error in the covariance matrix $\textbf{P}$. However, the number of available APs and hence, the localization method might change at any time, and we can use such information to improve the tracking performance with adaptive filtering. Since we have classified the arrangement of the available APs under six models (see Fig.~\ref{Models}), we can represent the adaptive filter as follows.  
The KF is implemented for each model as explained in Section~\ref{Kalman_Sub}, with the difference that $\textbf{E}_{\rm v}$ in \eqref{Ezz} is replaced with $\textbf{E}_{\rm v}^j$. The $\textbf{E}_{\rm v}^j$ is a $3\times3$ model-dependent covariance matrix  and can be expressed as follows
\begin{align}
   \textbf{E}_{\rm v}^j = {\rm diag}(\sigma_{\rm v}^2/\eta(j)) \,, \label{Ez}
\end{align}
where $\sigma_{\rm v}^2$ is the variance of the error in the location observation, and $\eta(j)$ is the adjustment parameter for the $j$th localization model. For models with high accuracy, we expect $\eta(j)$ to be high, which causes the diagonals of $\textbf{E}_{\rm v}^j$ to be low. This will result in a $\textbf{K}_t$ matrix with higher entries on the diagonal in \eqref{cov_state}, implying that a low measurement error is expected. For example, $\eta(5)$ should be high due to the expected high accuracy of Model-5, and $\eta(1)$ should be low due to the low accuracy of Model-1. Similarly, $\eta(0)$ should be set to a very small positive number to account for the high measurement error in Model-0. Note that by using a diagonal $\textbf{E}_{\rm v}^j$, we assume the estimation errors in different directions are uncorrelated as considered in the conventional KF model. 

Next, a final estimate for the state $\textbf{x}_t$ as in \eqref{kalman_gain}, and a covariance state $\textbf{P}_{t+1}'$ as in \eqref{cov_state} are obtained. Based on different covariance matrices $\textbf{E}_{\rm v}^j$ in \eqref{Ez}, the corresponding $\textbf{x}_t$ and $\textbf{P}_{t+1}'$ can be written as follows:
\begin{align}
    \textbf{x}_t = \tilde{\textbf{x}}_t^{(j)}\, \quad \rm{and} \quad  \textbf{P}_{t+1}' = \tilde{\textbf{P}}_{t+1}'^{(j)} \,, \label{model_dep}
\end{align}
where $\tilde{\textbf{x}}_t^{(j)}$ is the state estimate and, $\tilde{\textbf{P}}_{t+1}'^{(j)}$ is the state covariance matrix for Model-$j$ at time $t$.

\subsection{Coefficients of the Adaptive Filter}
Calculation of $\textbf{E}_{\rm v}^j$ for the adaptive KF is studied in~\cite{mohamed1999adaptive} and~\cite{yang2006optimal}. However, these studies do not assume certain estimation models as we do in this study. Instead, they calculate the optimal $\textbf{E}_{\rm v}^j$ or $\textbf{E}_{\rm x}^j$ matrices at each filter state with up-to-date parameters. Moreover, the ML estimation used in these studies has high computational complexity and may not be computed in real time without high computing power. Our problem differentiates from these studies by simplifying the adapting conditions to six distinct models unique to indoor VLC scenarios. We also simplify the problem of finding $\textbf{E}_{\rm v}^j$ to finding coefficients $\eta(j)$ in \eqref{Ez} by assuming the estimation errors in different directions are uncorrelated. 

\begin{figure}[tb]
	\centering
	\includegraphics[width = 3.25 in]{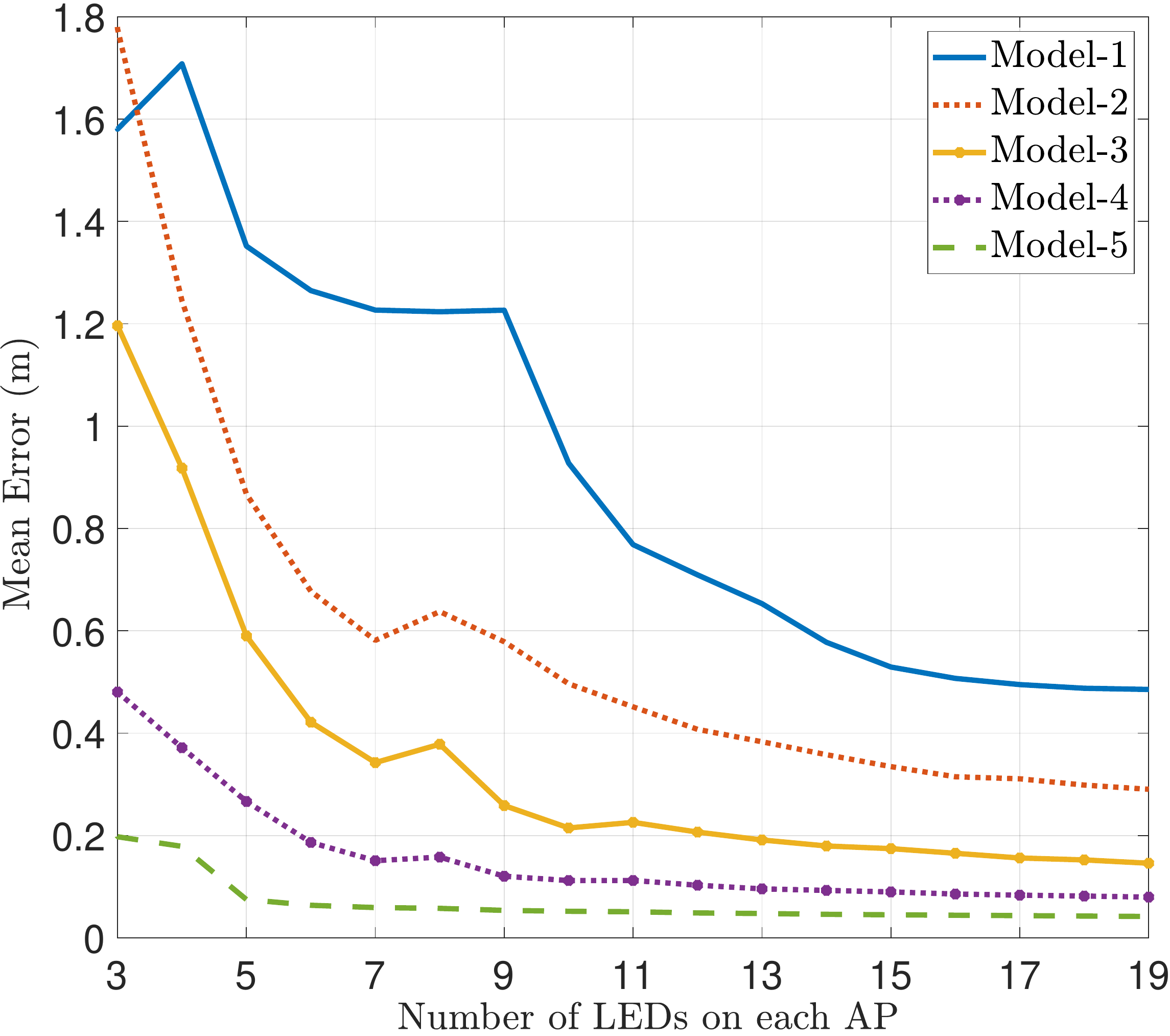}
	\vspace{-1mm}
	\caption{\small The mean localization error (without using KF) of each model for different number of LEDs per AP. }
	\label{main_room_error}
\end{figure}

While determining the coefficients $\eta(j)$, a good strategy would be to assign higher weights to models with lower localization errors. To examine this, we have simulated the mean localization error for each AP layout model in~Fig.~\ref{Models}. The mean errors are calculated by averaging the localization errors for different receiver locations. The results are presented in Fig.~\ref{main_room_error} for each model with a different number of LEDs on each AP and simulation parameters in Table~\ref{SimPar}. Model-0 is not included as this model does not provide any location estimation and relies only on prediction. The models with a higher number of accessible APs provide a lower error as expected. The only exception is that Model-1 performs better than Model-2 for the 3-LED case. This is because the ML estimation in \cite{Alphan_JLT} needs a large number of LEDs and APs to perform properly. While comparing the models with two available APs, we see that Model-3 performs better than Model-2. The reason for that is having both APs side by side in Model-2 may cause a bias in the estimation, i.e., the user can be estimated to be closer to the APs than the actual location. All models perform better when the number of LEDs on APs are higher, with a few exception data points. 

A similar problem with distinct layout models is studied in~\cite{8049459} for occupancy counting. In this study, the authors heuristically choose the coefficients such that each model has half the coefficient of the model with the next higher accuracy. Using a similar approach, we choose the coefficients of our model as follows:
\begin{align}
    \eta(0) = \frac{1}{32}, \quad \eta(1) = \frac{1}{16}, \quad \eta(2) = \frac{1}{8}, \label{fixedCoeff} \\
    \eta(3) = \frac{1}{4}, \quad \eta(4) = \frac{1}{2}, \quad \eta(5) = 1. \quad  \nonumber
\end{align}
The coefficient of the Model-5 is set to the highest value 1 because all the APs are available to the user in this model. Besides, setting $\eta(5)$ to 1 provides a fair comparison with the conventional KF. 
Since the coefficients in~\eqref{fixedCoeff} do not depend on the exact expected error of the models, we will refer to them as the \textit{fixed} coefficients in the rest of the paper. 

The fixed coefficients in \eqref{fixedCoeff} do not use all the information about how accurate each model is. Since we know the expected errors for each model, we can use this information to calculate better $\eta(j)$ coefficients. With this purpose, we propose a heuristic method, where the coefficient of each model is inversely proportional to the expected error of the model. Let $\omega(j, N)$ denote the mean location error for model $j$ and $N$ LEDs per AP and $\tilde{\eta}(j, N)$ denote the coefficient $\eta(j)$ for $N$-LED case. Then, we have
\begin{align}
    \tilde{\eta}(j, N) = \omega(5, N)/\omega(j, N). \label{coeff}
\end{align}
Note that using~\eqref{coeff}, we get $\tilde{\eta}(5, N)=1$ for all $N$. 

In Fig.~\ref{main_room_error}, we observe that the expected errors saturate as the number of LEDs increase. We can consider these saturation errors as the lower bound of each model. Therefore, as a third option, we can evaluate each model at the best performance and determine the adaptive coefficients as 
\begin{align}
    \eta_\infty(j) = \tilde{\eta}(j, \infty). \label{coeffInf}
\end{align}
In the simulations, we will find the coefficients using \eqref{fixedCoeff}, \eqref{coeff}, and \eqref{coeffInf}, separately, and compare their performances to investigate the sensitivity of adaptive KF to different coefficients. The coefficients in \eqref{coeff} and \eqref{coeffInf} are calculated using the data given in Fig.~\ref{main_room_error}. 

\section{Numerical Results}\label{sec:results}

\begin{table}
	\caption{Simulation Parameters.}
		\vspace{-4mm}
	\begin{center}
		\begin{tabular}{ |c|c| }
			\hline
			LED directivity index, $\mode$ & 10 \\
			\hline
			The vertical tilt angle of LEDs, $\alpha$ & 25$^\circ$ \\
			\hline
			Additional vertical tilt angle of LEDs, $\beta$ & 10$^\circ$ \\
			\hline
			Receiver FOV, $\FOV$ & 25$^\circ$ \\
			\hline
			Effective surface area of a PD, $A$ & 1 cm$^2$ \\
			\hline
			Room size & 6$\times$6$\times$3 m$^3$ \\
			\hline
			Default device height, $\nu$ & 0.9 m \\
			\hline
			The standard deviation of estimation error, $\sigma_x$ & 0.005 \\
			\hline
			The standard deviation of measurement error, $\sigma_y$ & 0.05 \\
			\hline
		\end{tabular}
		\label{SimPar}
	\end{center}
	\vspace{-1mm}
\end{table}

In the simulations, we assume that a user walks along a random waypoint path. The user starts walking from one side of the room and visits the randomly chosen waypoints. The height of the waypoints is also random with $\mathcal{U}(0.7~m, 1.1~m)$. The waypoints are chosen so that the turn angle cannot exceed 90$^\circ$. If this condition is not satisfied, the waypoint is randomly selected again until it is satisfied. The path ends at the maximum length of 30~m, or when new waypoint selection is not possible. The user speed is 0.1~m per time step, and therefore the VLC user's location is estimated at each 0.1~m. Other simulation parameters are provided in Table~\ref{SimPar}. The simulations are repeated for 2000 routes and average results are provided. \looseness=-1

\begin{figure}[t]
	\centering
	\includegraphics[width = 3.25 in]{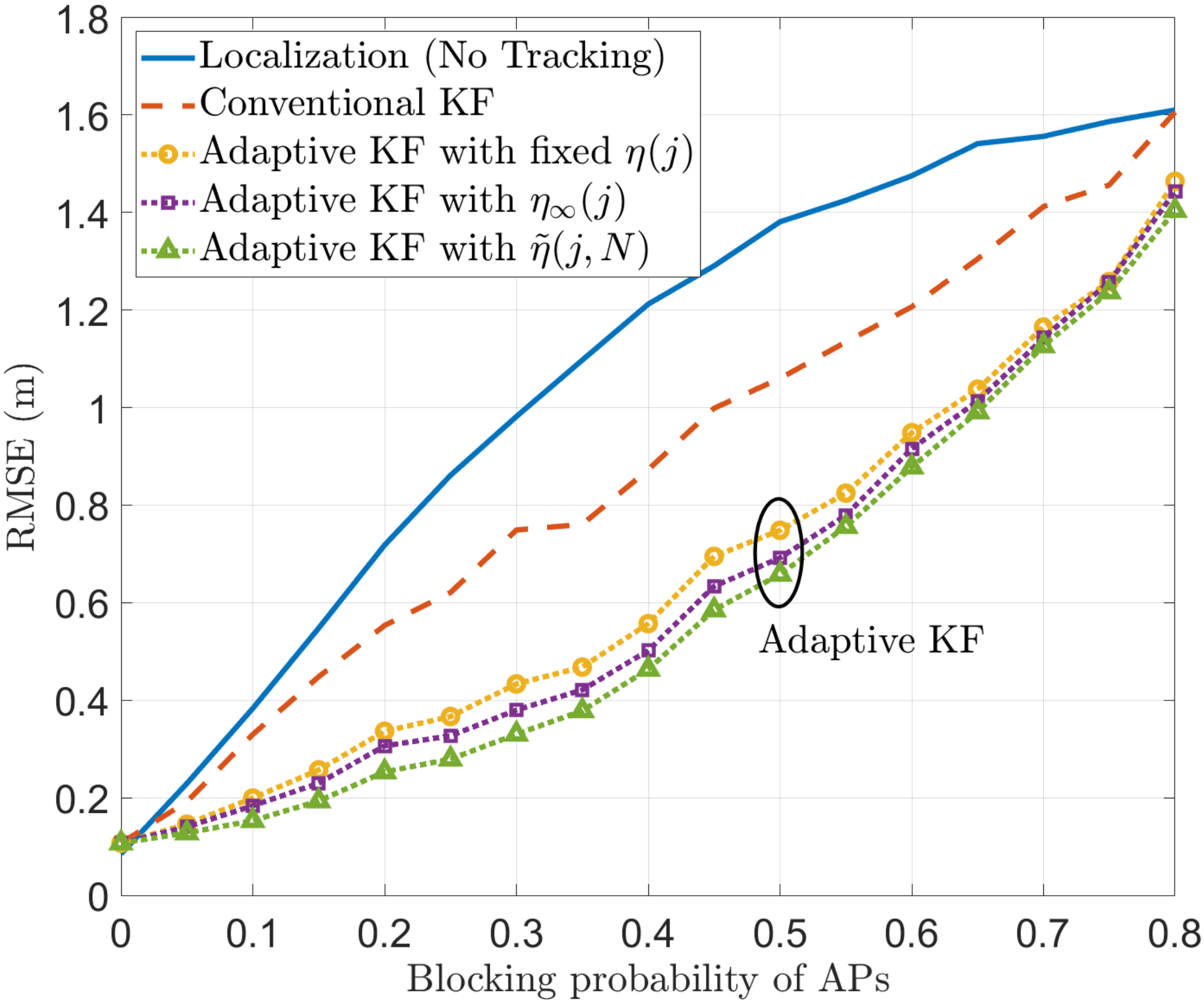}
	\vspace{-2mm}
	\caption{\small The RMSE for varying probability of blockage for APs, while each AP has seven LEDs as shown in Fig.~\ref{7Element}. }
	\vspace{-1mm}
	\label{Varying_Block_Prob}
\end{figure}

The APs are assumed to be blocked with a random probability at each time instant. This is a scenario where the available APs change frequently (e.g., due to random receiver orientation~\cite{eroglu_TCOM}) and is chosen to show at what extent the adaptive filter can improve the accuracy. Fig.~\ref{Varying_Block_Prob} shows the RMSE for localization over the path for increasing probability of each AP being blocked. There are seven LEDs on each AP with the configuration shown in Fig.~\ref{7Element}. The solid line shows the RMSE for unfiltered localization, the dashed line shows the RMSE for the localization with the conventional KF, and the dotted lines show the RMSE with the adaptive KF. 

For the results in Fig.~\ref{Varying_Block_Prob}, the layout model is decided based on the available APs at each time step as shown in Fig.~\ref{Models}. When the blocking probability is equal to zero, the layout model is always Model-5. In this case, the conventional and adaptive filters are identical, and they do not provide a visible improvement on the accuracy over the unfiltered localization. The reason for that is, the error is already low, and it is not necessarily Gaussian distributed, so the KF does not necessarily provide improvement. When the blocking probabilities increase, the RMSE increases too. In this case, the conventional KF improves accuracy significantly. The adaptive KF reduces the RMSE even more by utilizing the layout model information. Using the heuristic in \eqref{coeff} or \eqref{coeffInf} provides additional gain compared to using fixed coefficients. The coefficients in \eqref{coeff} are $\tilde{\eta}(j,7)$, which considers the mean error for 7-LED case. These coefficients yield lower error than $\eta_\infty(j)$, which shows that the proposed heuristic utilizes the expected error information accurately. \looseness=-1

\begin{figure}[t]
	\centering
	\includegraphics[width = 3.25 in]{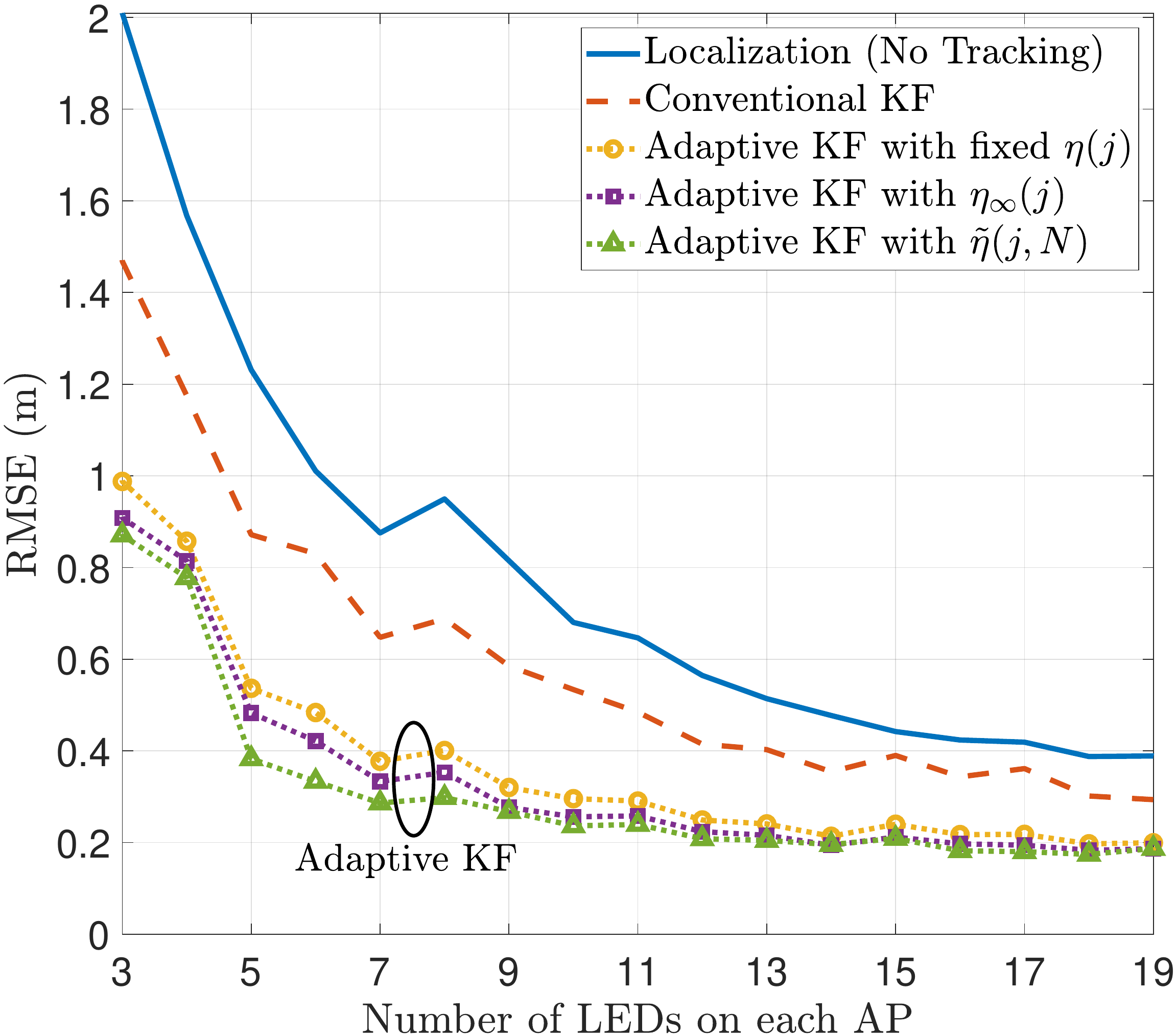}
	\vspace{-1mm}
	\caption{\small The RMSE for varying number of LEDs per AP. The blocking probability of each AP is set to 0.25. }
	\vspace{-5mm}
	\label{Varying_LED_number_RMSE}
\end{figure}

Fig.~\ref{Varying_LED_number_RMSE} shows the RMSE for varying number of LEDs per AP. The receiver is again assumed to move along the random waypoint path, and the blocking probability of any AP is 0.25. When the number of LEDs increases, the RMSE steadily decreases due to larger number of measurements that can be used in the hybrid localization. The only exception is the 8-LED case, where there is a single LED at the outermost layer (see Fig.~\ref{19Element}), which might be causing bias in the estimation. This is also visible in Fig.~\ref{main_room_error} for Model-2 to Model-4, where the 8-LED case performs worse than the 7-LED case. In Fig.~\ref{Varying_LED_number_RMSE}, we see that the adaptive KF provides a significant improvement over the conventional KF for any number of LEDs. Moreover, with the adaptive KF, the RMSE saturates faster for increasing number of LEDs compared to using the conventional KF. 
Therefore, the adaptive KF provides higher accuracy with lower number of LEDs, removing the need for high number of LEDs to achieve higher accuracies. Overall, the RMSE of adaptive KF is about 30\% to 50\% of unfiltered localization RMSE for any number of LEDs. Using the coefficients $\tilde{\eta}(j,N)$ provides the least RMSE in general. As \textit{N} goes to 20, $\tilde{\eta}(j,N)$ and $\eta_\infty(j)$ becomes identical, and they yield the same RMSE. 

\vspace{-2mm}
\section{Conclusion}\label{sec:conclusion} 
In this study, we propose the use of adaptive KF for multi-element VLP systems with intermittent AP availability. The VLP heavily relies on LOS links, in whose absence, an AP could be completely inaccessible. Localization accuracy may suffer from the frequent changes in the availability of APs to the user. The adaptive KF takes this into account and makes a final estimation considering the different AP layout models, localization methods, and their expected accuracies. To evaluate the adaptive KF, we consider a scenario where APs can be randomly blocked, and choose our localization methods depending on the layout of the accessible APs. We use simulations to decide the localization accuracy in each layout model, and as a heuristic, we choose the adaptive coefficient of each model inversely related to their expected localization error. The simulation results show that the adaptive implementation improves the performance of the KF significantly and yields much lower RMSE in all scenarios. \looseness=-1

\vspace{-1mm}
\bibliographystyle{IEEEtran} 
\bibliography{IEEEabrv,mypaper}
\end{document}